# Spin and Orbital Magnetism in UH$_2$ Thin Films Studied by X-ray Magnetic Circular Dichroism

Evgenia A. Tereshina-Chitrova[1], Mykhaylo Paukov[2], Oleksandra Koloskova[1], Amir Hen[3], Fabrice Wilhelm[3], Lukas Horak[4], Mayerling Martinez Celis[4,5], Miroslav Cieslar[4], Ladislav Havela[4], Andrei Rogalev[3], Thomas Gouder[6]

[1] *Institute of Physics, Czech Academy of Sciences (FZU), Prague, Czech Republic*

[2] *Centrum výzkumu Řež, Hlavní 130, 250 68 Husinec-Řež, Czech Republic*

[3] *European Synchrotron Radiation Facility (ESRF), Grenoble, France*

[4] *Faculty of Mathematics and Physics, Charles University, Prague, Czech Republic*

[5] *Normandie Univ, UNICAEN, ENSICAEN, CNRS, Laboratoire CRISMAT, 6 Bvd du Maréchal Juin, 14050, Caen, France*

[6] *European Commission, Joint Research Centre (JRC), Postfach 2340, D-76125, Karlsruhe, Germany*

Uranium dihydride UH$_2$ is a metastable phase unknown in bulk form but accessible through thin-film synthesis. We prepared UH$_2$ films by reactive *dc* sputtering on CaF$_2$(001) or Si(001) substrates, the latter equipped with a Mo buffer layer to suppress a U-Si interdiffusion. On CaF$_2$, UH$_2$ adopts the fluorite-type structure with a near-[1 1 1] out-of-plane texture, four rotational domains, and a lattice parameter $a$ = 539 ± 3 pm without measurable strain, whereas the Mo-buffered film is polycrystalline. X-ray photoelectron spectroscopy confirmed complete hydrogenation and minimal oxidation. Magnetization and XMCD measurements show ferromagnetic ordering with Curie temperatures of 120-130 K and a uranium 5$f$ moment of ≈ 0.9 $\mu_B$/U, dominated by the orbital contribution ($\mu_L$ ≈ 1.4 $\mu_B$, $\mu_S$ ≈ −0.5 $\mu_B$), in a good agreement with GGA+$U$ computations, which otherwise overestimate absolute values of the spin and orbital components. The slightly reduced moment in thinner CaF$_2$-supported films is attributed to surface U(IV) species. These results demonstrate that thin-film synthesis enables stabilization of UH$_2$ and direct probing of 5$f$ magnetism, opening pathways toward higher uranium hydrides and interface-engineered actinide systems.

**Introduction:**

In the realm of nuclear safety, the corrosion of uranium by hydrogen—which may be released from moisture by radiolysis or arise from other environmental sources—poses a significant hazard [1]. The reaction forms pyrophoric uranium hydrides, which are highly reactive powders that complicate handling in non-specialized laboratories. Consequently, safe experimentation is typically only possible using small, localized hydride regions on metallic uranium surfaces.

Beyond their safety implications, uranium hydrides have garnered growing interest due to their exceptional functional properties, particularly their fully reversible hydrogen sorption [2]. While their high atomic mass precludes mobile applications, uranium hydrides excel in stationary systems owing to their high volumetric hydrogen density (≈130 g cm$^{-3}$, nearly twice that of liquid hydrogen) and wide tunability of H$_2$ equilibrium pressure (10$^{-2}$ Pa at 300 K to 10$^5$ Pa at 700 K) [3]. These attributes, combined with robustness and cost-effectiveness, make uranium hydrides promising for tritium handling in fusion devices [4], hydrogen purification [5], and integration into nuclear fuel cycles [6].

From the fundamental perspective, uranium hydrides present a rare case, exhibiting both unusual formation behavior and remarkable magnetic properties. Whereas most 4$f$ and 5$f$ metals form both dihydride and trihydride phases [7], uranium under equilibrium conditions yields the trihydride UH$_3$ [8], known in two polymorphs: metastable α-UH$_3$ (cubic, $Pm$-$3n$, $a$ = 416 pm) and stable β-UH$_3$ (more complex cubic structure, $a$ = 664.4 pm). Both exhibit ferromagnetism with Curie temperatures $T_C$ of ≈170 K [8,9], remarkably high for an $f$-metal hydride. The dihydride phase UH$_2$ does not form under equilibrium conditions and had never been synthesized in bulk. However, it was synthesized in a thin-film form [10], adopting the $fcc$ CaF$_2$ structure ($a$ = 536 pm). It exhibits ferromagnetic behaviour with a lower $T_C$ ≈ 120 K [10,11]. This marked a significant step forward in uranium hydride research, opening access to a previously unobserved phase and its distinct magnetic characteristics.

The magnetism of U hydrides challenges the expectations set by the Hill limit rule [12]. Despite the largest U–U spacing (≈378 pm) among uranium hydrides, UH$_2$ exhibits a lower $T_C$ and a reduced magnetic moment (previously estimated from bulk magnetization as ≈ 0.5 μ$_B$/U [11], 0.89 μ$_B$/U from theory [13]) compared to both variants of UH$_3$, which have significantly lower the shortest U–U distances (360 pm in α-UH$_3$, 332 pm in β-UH$_3$) and magnetic moments, close to 1 μ$_B$/U. This contradicts the traditional band concept, which links larger U-U spacing to more localized 5$f$ states and stronger magnetism. Theoretical works [13,14] instead suggest that the U-H interaction (mainly 6$d$-1$s$) is more important than the U-U separation in governing the 5*f* magnetism in uranium hydrides, highlighting the need for direct experimental verification. Because magnetic moments derived from macroscopic measurements on thin films suffer from large uncertainties, we employed X-ray magnetic circular dichroism (XMCD) spectroscopy at the uranium $M_{4,5}$ edges, which provides not only total 5*f* moments but also the orbital and spin components via magneto-optical sum rules [15]. By confronting these XMCD-derived quantities with various ab-initio calculations, we aim to resolve the underlying mechanism of 5*f* magnetism in uranium hydrides, especially considering the possibility that substrate-induced effects in thin films may further modulate electronic and magnetic properties.

In this work we present the synthesis and comprehensive characterization of $UH_2$ thin films grown on two types of substrates, $CaF_2$ (001) and Si (001), chosen to explore possible substrate-induced effects on structure and magnetic behaviour. $CaF_2$ was selected due to its fluorite structure with the lattice parameter similar to $UH_2$, while Si allows comparison with previously reported $UH_2$ thin films [10]. The primary objectives of this study are (1) to establish controlled synthesis conditions for $UH_2$ thin films and investigate their structural properties; (2) to characterize their magnetic behaviour using magnetometry and XMCD, and compare the results for $UH_2$ films exhibiting different structural arrangements; and (3) to correlate experimental findings with available theoretical predictions [13,14] to provide a comprehensive understanding of the structure-property relationships in uranium dihydride thin films.

**Experimental details:**

$UH_2$ thin films with controlled impurity levels (Fig. 1) were prepared by reactive DC sputter deposition using a natural U rod as a target (99.9 wt.% purity, 5 mm diameter). The deposition was carried out in an Ar–$H_2$ (hydrogen of 5N purity, further purified using an Oxysorb® cartridge) atmosphere using an in-house triode sputtering system [11], with a base pressure of $10^{-9}$ Pa. The hydrogen partial pressure required for full hydride formation was determined experimentally (Fig. 2). Plasma was maintained using a thoriated tungsten filament, which emitted electrons to sustain the discharge. The target voltage was −700 V, with the current controlled by adjusting filament emission parameters. The film thickness scales approximately with deposition time and target current, allowing a rough estimate of the growth rate under the present conditions. For the range of parameters used in this study (target current 2–3 mA, deposition time 15–400 min), the resulting $UH_2$ film thicknesses are on the order of several tens to a few hundred nanometers.

We used two types of single crystalline substrates, CaF$_2$ with the [001] orientation and Si(001) wafers. The CaF$_2$ substrates were degassed by annealing at 673 K for 1200 s. The films grown on CaF$_2$ were deposited at room temperature to promote the formation of the metastable UH$_2$ phase and to minimize thermal stress on the brittle substrate. To avoid excessive self-heating of the hydride during sputtering, the deposition time was limited to 1000 s, consistent with previous reports [11]. The Si(001) wafers used as a substrate were cleaned at 673 K by bombardment of Ar$^+$ ions accelerated to 2 keV energy. The deposition onto Si substrates was carried out for 4000 s with the substrate cooled to 177 K and with a Mo buffer layer applied, in order to suppress the reaction between U and Si that would otherwise lead to the formation of USi$_3$ as known from previous studies [16]. After deposition, both films were capped with a few nanometer thick Mo layer to prevent oxidation and environmental contamination.

After deposition, the samples were transferred under UHV conditions (transfer time < 3 min) from the growth chamber to the analysis chamber for immediate surface analysis prior to capping. X-ray photoelectron spectroscopy (XPS) was performed using monochromated Al K$\alpha$ radiation (1486.6 eV) and a SPECS PHOIBOS 150 MCD-9 hemispherical analyzer, calibrated to the Au 4$f_{7/2}$ (83.9 eV) and Cu 2$p_{3/2}$ (932.7 eV) lines of pure metals.

For ex-situ structural characterization, the films were examined using a Rigaku SmartLab diffractometer equipped with a 9 kW Cu rotating anode source. The primary beam of Cu-K$\alpha$ radiation ($\lambda$ = 0.15418 nm) was parallelized in the equatorial direction by an X-ray parabolic mirror. For texture determination, an in-plane pole figure measurement technique was used, where the axial divergence and acceptance were limited to 5° by Soller slits, and the equatorial acceptance was defined by a parallel slit analyzer to 0.5°. Wide Reciprocal Space Maps (WRSM) were measured in low-resolution mode with a 15 cm sample-to-detector distance, using a small beam collimated in both directions to 0.5 × 0.5 cm$^2$ and a 2D single-photon-counting hybrid detector without any optics in front. Mapping of intensity distribution in vertical planar cuts of reciprocal space was performed for nine sample azimuths, separated by an azimuthal step of 15°.

Transmission electron microscopy (TEM) was used to determine film thickness and assess elemental and phase composition. The measurements were performed on a JEOL JEM-2200FS microscope operating at 200 kV. Cross-sectional TEM lamellae were prepared using a ZEISS Auriga Compact scanning electron microscope (SEM) equipped with a focused ion beam (FIB) system and an in-situ EasyLift manipulator for site-specific lift-out.

Magnetic measurements were carried out using a Quantum Design PPMS 9 system equipped with a vibrating sample magnetometer (VSM) for enhanced sensitivity. Electrical resistivity, $\rho(T)$, was measured using the Van der Pauw method. The films were contacted using 30 µm-thick Al wires containing 3% Si, bonded to the sample surface by means of a wire bonder. In both magnetization and transport experiments, an external magnetic field was applied in-plane, parallel to the film surface.

X-ray absorption spectroscopy (XAS) and X-ray magnetic circular dichroism (XMCD) experiments were carried out at the ESRF beamline ID12, which is dedicated to polarization-dependent X-ray absorption spectroscopy studies in the photon energy range from 2 to 15 keV. All measurements were performed at the ID12 high-field XMCD end-station based on a 17 T superconducting solenoid [17]. For the experiments at the U-$M_4$ (≈3.727 keV) and $M_5$ (≈3.552 keV) edges, the source was a Helios-II type helical undulator providing a high flux of circularly polarized X-ray photons. After monochromatization with a double crystal Si(111) monochromator, the degree of circular polarization was reduced to ≈0.44 at the $M_4$ edge and ≈0.35 at the $M_5$ edge.

The XAS spectra at both the U-$M_{4,5}$ edges were recorded in the total fluorescence yield (TFY) detection mode in backscattering geometry, for parallel $\sigma^+(E)$ and antiparallel $\sigma^-(E)$ alignments of the photon helicity with respect to the external magnetic field (+/- 17 T) applied along the beam direction. The spectra were then corrected for self-absorption effects. The corrected spectra are nearly identical to those recorded using the total electron yield (TEY) method, confirming the validity of the applied corrections (see Supplementary information).

Due to the large difference in the binding energies of the uranium core $3d_{3/2}$ and $3d_{5/2}$ states (176 eV), the $M_4$ and $M_5$ spectra were measured separately, and a proper normalization procedure was applied. To avoid an arbitrary choice in spectra normalization and in the energy position of the step function simulating the transitions into the continuum, we adopted the method described in [15]. This method is based on the assumption that the EXAFS oscillations are identical above the two spin-orbit split edges, i.e. $M_5$ and $M_4$, for isotropic XAS spectra recorded under the same experimental conditions. Accordingly, the spectra recorded at the $M_4$ and $M_5$ edges were shifted to a common energy scale and normalized such that the first EXAFS wiggles at both edges overlap. The U edge jump intensity ratio $M_5/M_4$ was then fixed to 6:4 consistent with the statistical occupation numbers of the two spin-orbit split core levels with $j = 3/2$ and $j = 5/2$.

The XMCD spectra were obtained as the direct difference between the normalized and self-absorption corrected spectra recorded with left and right circularly polarized X-rays, taking into account the incomplete degree of circular polarization. The magnetization curves were measured by monitoring the XMCD intensity at the Uranium $M_4$-edge as a function of the applied magnetic field from +17 T to -17 T.

**Results and discussion.**

1. **In-situ XPS study: formation of hydrides**

XPS was employed to evaluate the chemical composition to detect possible impurities (Fig. 1), as well as to monitor the formation of uranium hydride during sputter deposition in the Ar/$H_2$ atmosphere (Fig. 2). In both the $CaF_2$-based and Si-based samples, the spectra show only a minor oxygen contribution of no more than 2 at.% (cf. the supplementary XPS data in Ref. [18]), which is attributed to surface oxidation, produced by the tendency of $UO_2$ to segregate at the surface of metallic systems. A weak but distinct fluorine signal (F-1$s$ peak at ≈ 685 eV), which can be quantified as 5 at.% F, has been detected at the surface of the $CaF_2$-supported film, while Ca contamination is below detection threshold (no Ca-2$p$ is observed). Considering the ≈40 nm thickness of the $UH_2$ film (see details in Section 2) and the attenuation length of photoelectrons with 800 eV kinetic energy, which does not exceed 1.5 nm, the signal cannot originate from the substrate. Instead it can be taken as indication that a small amount of substrate fluorine has been embedded into the U-H film due to the intermixing at the interface and segregated at the surface. As we can assume the chemical form $UF_4$ or $UO_2F_2$, both being insulating, they do not contribute the electronic states near the Fermi level, $E_F$, or to electrical transport. No F-related features are observed in the Si-based $UH_2$ film (cf. Fig. 1).

Gradual addition of hydrogen into the sputter gas causes systematic and reproducible changes in the U-4$f$ core-level spectra (Fig. 2). Although the hydrogen content cannot be directly quantified by XPS since the H-1$s$ states are part of the valence band [11], the evolution of the U-4$f$ spectral shape, namely the broadening into a triangular shape, provides a reliable qualitative indicator of hydride formation when reaching a stable shape beyond a certain threshold pressure. This saturation behaviour, typically occurring at $H_2$ concentrations around 5% in Ar ($p_{H2}$ ≈ 4 × 10$^{-2}$ Pa at $p_{(Ar+H2)}$ = 0.83 Pa total in Fig. 2), indicates the formation of a homogeneous hydride phase rather than a mixture of a metal and sub-hydride phases [11]. For production of our $UH_2$ sample, we used a higher $H_2$ partial pressure of ≈ 6x10$^{-2}$ Pa (7.5% $H_2$ in Ar), at the same total pressure, to provide a safe margin. We also observed a small shift of approximately 0.2 eV in the 4$f$ peaks maxima upon hydride formation (Fig. 2), compared to pure U metal, where the 4$f_{5/2}$ and 4$f_{7/2}$ peaks are located at 388.2 eV and 377.3 eV binding eneregy (BE), respectively. This shift can be attributed to a reduction in the correlation energy between the 4$f$ core level and the outer electrons, which are partially transferred from the U atom due to chemical bonding with hydrogen [19]. The asymmetric broadening of the 4*f* lines is due to lattice expansion accompanying hydride formation, bringing higher density of states at $E_F$ and stronger e-e correlations [10].

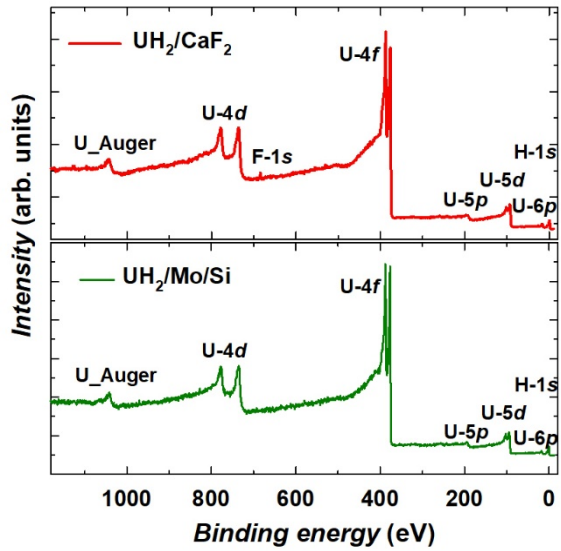 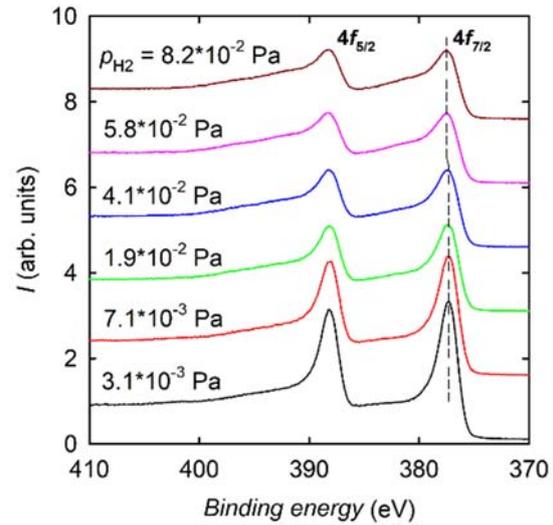

Fig. 1. Overview XPS spectra of $UH_2$ films deposited on $CaF_2$ (top) and on a Si substrate with a Mo buffer layer (bottom), acquired prior to Mo capping.

Fig. 2. Evolution of the U-4*f* core-level spectra with increasing $H_2$ partial pressure in an Ar + $H_2$ atmosphere at a total pressure of 0.8 Pa.

While XPS demonstrates complete hydrogenation of uranium, it does not provide direct information about the H/U ratio in the sample. Hence, to distinguish between the two possible hydride phases, $UH_2$ and $UH_3$, we have to rely on structural and magnetic data presented below.

## 2. Crystal structure study

The choice of substrate strongly influences the structural characteristics of $UH_2$ films, particularly their preferred orientation and strain state. For $UH_2$ films grown on Si substrates with a Mo buffer, a polycrystalline $CaF_2$-type structure ($a \approx 536$ pm) and residual compressive strain of about –1.5 GPa have been reported [10]. The Si substrate plays no role, as the film texture is governed by the Mo buffer layer and the sputter deposition dynamics. In contrast, $UH_2$ films grown on $CaF_2$(001) exhibit a highly specific crystallographic texture, as revealed by pole figure analysis (Fig. 3 (c)). Instead of growing coherently with the substrate lattice in [001] direction, the films preferentially adopt an out-of-plane near-[1 1 1] orientation, i.e. (1 1 1)$_{UH_2}$ almost parallel to (0 0 1)$_{CaF_2}$, with the exact in-plane alignment of [0 1 -1]$_{UH_2}$ // [1 -1 0]$_{CaF_2}$ (see Fig. 4). Following the four-fold symmetry of the $CaF_2$ surface, four equivalent rotational domains are formed, each having [0 1 -1]$_{UH_2}$ aligned with one of the substrate directions [1 1 0], [1 -1 0], [-1 -1 0], and [-1 1 0]. The misorientation from the ideal [111] texture in these domains is expressed by the tilt of the [1 1 1] UH$_2$ axis that deviates by 12.4 ± 0.7° toward its respective in-plane substrate direction [±1 ±1 0]$_{CaF_2}$, while the in-plane alignment of [0 1 -1]$_{UH_2}$ is preserved as the rotation axis of the tilt (see visualisations in Fig. 3 (b) and Fig. 4).

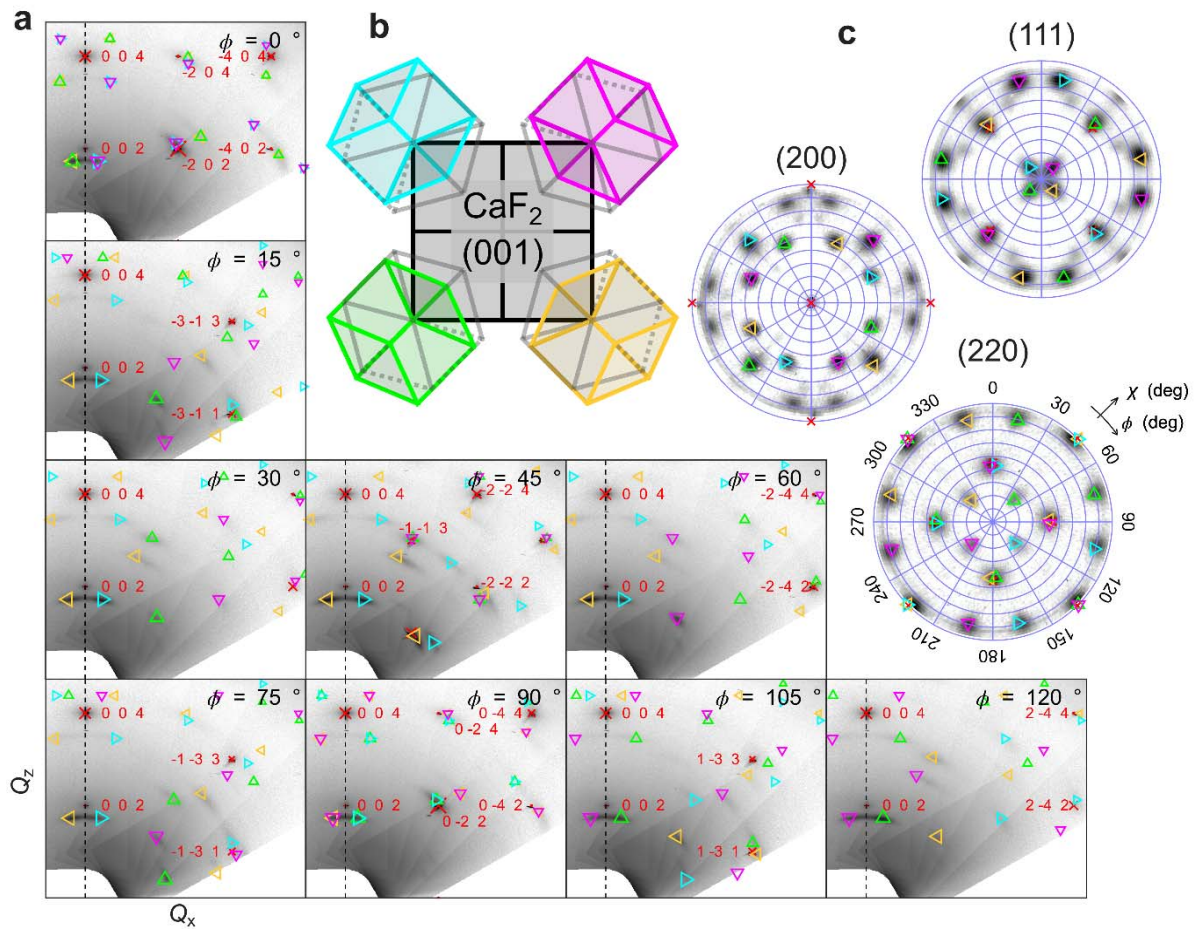

Fig. 3. (a) Low-resolution reciprocal space maps (RSMs) of the UH$_2$ film grown on CaF$_2$(001), recorded at different azimuthal angles φ, showing the diffracted intensity distribution in vertical cuts of reciprocal space. The dashed line indicates the $Q_z$ axis. (c) Schematic of the four texture components considered in the XRD analysis. Colours indicate the tilt directions, matching those used in the (a) and (c) figures. (c) Pole figures of the UH$_2$ film for the (1 1 1), (2 0 0), and (2 2 0) reflections. Substrate poles are marked by red crosses, while UH$_2$ reflections are shown as coloured triangles corresponding to four crystallographic domain orientations. Numerical simulations based on the structural model described in the text are overlaid on both RSM and pole figure data.

Unlike the polycrystalline $UH_2$, the $UH_2/CaF_2(001)$ film shows no clear evidence of residual strain. Reciprocal space mapping (RSM, Fig. 3) yields a lattice parameter of 539 ± 3 pm; any remaining strain is within the experimental uncertainty. This difference in crystal structure between fully polycrystalline $UH_2$ and the epitaxially aligned (multi-domain) $UH_2$ film on $CaF_2(001)$ may influence hydrogen uptake and, consequently, its magnetic properties. The microstructure of the latter film is examined in more detail by TEM below. A schematic of the orientation relationship with the substrate is shown in Fig. 4.

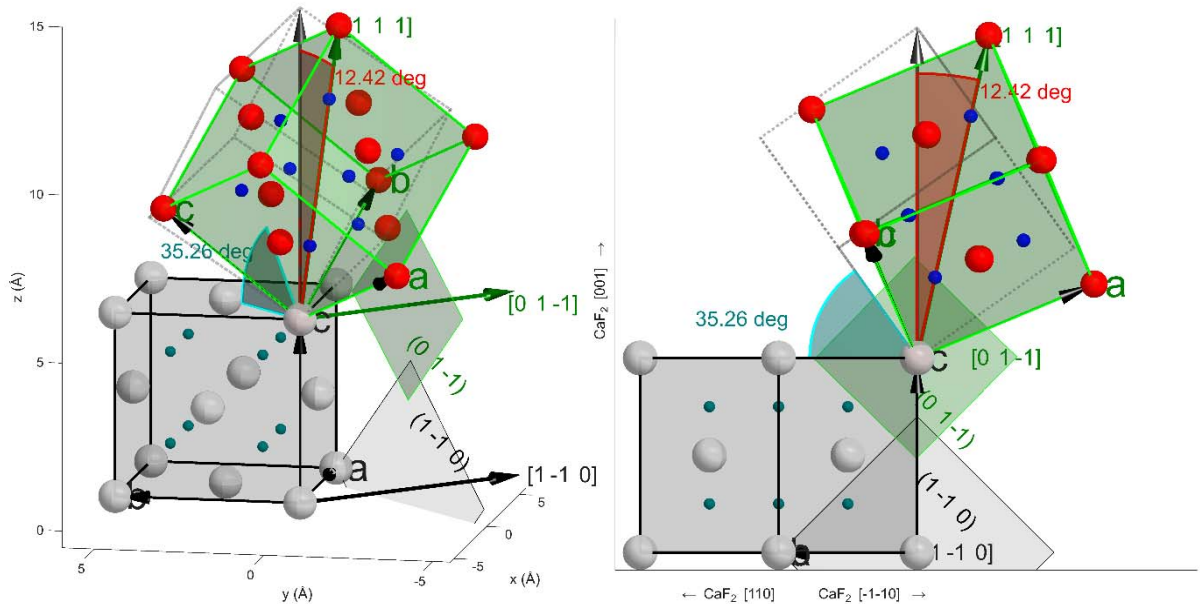

Fig. 4. Spatial visualisation of the orientation model of the $UH_2$ film on $CaF_2(001)$. The film consists of four rotational domains, each with the [1 1 1] axis tilted by ≈12.4° from the surface (substrate) normal. Direction [0 1 -1]$_{UH_2}$ remains in the surface plane, so that (0 1 -1)$_{UH_2}$ is parallel to (1 -1 0)$_{CaF_2}$, both having similar d-spacings.

It is important to note that actinide and rare-earth dihydrides of this structure type often exhibit non-stoichiometric compositions, typically described as $RH_{2+x}$, with $x$ reaching up to 0.75 [20,21]. This corresponds to additional randomly occupied octahedral hydrogen sites (the native H sites in $RH_2$ are tetrahedral), introducing a degree of structural disorder into the system. Such variability in hydrogen concentration may impact physical properties of the films, including magnetism. In fact, a slight difference in Curie temperature was observed between the films grown on $CaF_2$ and those on Si substrates (see next Section). The film deposited on $CaF_2$ exhibited a somewhat higher $T_C$, which may reflect not only subtle strain differences but also variations in the hydrogen-to-uranium ratio, which are not detectable by XPS or XRD.

This observation highlights the sensitivity of the UH$_2$ system to both growth conditions and local stoichiometry.

In order to determine the microstructure of the sputter deposited hydrides and their elemental mapping, TEM studies were performed on the cross-section of selected thin films. Fig. 5 shows a global view of the UH$_2$ thin film deposited on the CaF$_2$ substrate, the Mo cap, and the Pt protection, used to avoid an excessive damage of the very surface by Ga$^+$ ions during the lamella fabrication. The elemental mapping (not shown here), performed in a scanning TEM mode, indicates a lack of oxygen contamination, hence we assume the presence of the U hydride only. The issue is to determine which hydride is present.

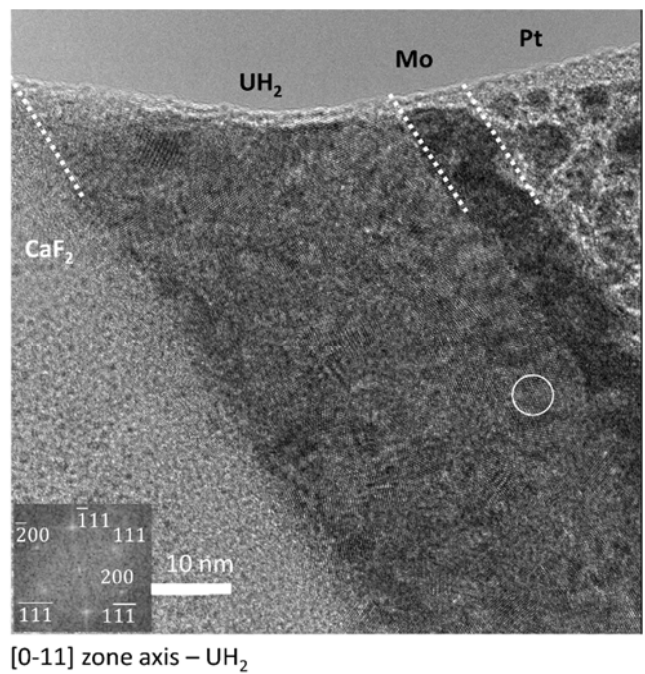

Fig. 5. TEM micrograph of the UH$_2$ film deposited onto the CaF$_2$ substrate and FFT pattern of the regions highlighted by white circle in the film (inset at the bottom left).

The analysis of several regions using the Fast Fourier Transform (FFT) method points to the CaF$_2$ structure type of UH$_2$. The orientation can be identified using a FFT analysis, yielding patterns of particular crystal directions. The most common orientation corresponds to the [1 -1 0] orientation perpendicular to the image plane and the [1 1 1] direction close to the growth direction. An example illustrating it is in Fig. 5, in which the analysed area is marked by the white circle, and the FFT pattern is at the bottom left. This reveals a texture, which is actually more pronounced at the top of the film than next to the substrate. We can conclude that although the UH$_2$/CaF$_2$ film exhibits a strong structural relationship with the substrate,

TEM analysis reveals that it is better described as polycrystalline with a pronounced [1 1 1] texture rather than epitaxial.

The total thickness of the $UH_2$ film in the place where the lamella was extracted from is ≈ 40 nm. The grain size determination performed in several grains indicated an average grain size of ≈ 7 nm with grains roughly equiaxial. The lattice parameter can be estimated as $a ≈ 533$ pm. Slight decrease of $a$ would again indicate increasing H occupancy in all $RH_{2+x}$; however, the difference observed by XRD and TEM in films are within the experimental error.

## 3. Magnetization and electrical resistivity data

Magnetization measurements of the $UH_2$ films were performed with the external magnetic field applied in the plane of the film (Fig. 5). The full magnetic saturation was not reached within the maximum applied field of 9 T, and the measurements remained in the minority loop regime. Nevertheless, we made a very approximate estimate of magnetic moment per U based on the nominal film thickness and area, yielding $\mu_U ≈ 0.5\text{-}0.6\ \mu_B$. Theoretical studies [13] indicate that the [1 1 1] direction is the easy-magnetization axis in $UH_2$. In our $CaF_2$-grown films, which exhibit a [1 1 1] out-of-plane texture, the applied in-plane magnetic field is therefore not aligned with the easy magnetization axis. Typically, shape anisotropy in thin films favors in-plane magnetization due to demagnetizing fields, and it can dominate over the magnetocrystalline anisotropy when the latter is modest. However, uranium hydrides are known to exhibit large magnetocrystalline anisotropy [10,11], which may counteract the shape-induced tendency and maintain a strong out-of-plane [1 1 1] magnetic preference. This misalignment between the field and the easy-axis directions, along with high anisotropy energy barriers, likely reduces the moments and contributes to the incomplete saturation and minor loop behaviour observed in the $M(H)$ data, even under fields as high as 9 T.

When comparing the two films on different substrates, one notices a slightly higher $T_C$ for the film grown on $CaF_2$ (130 K) compared to that on Si (120 K) (see Fig. 5). While strain could play a role in the Si-based sample, another possible explanation involves substrate-induced differences in hydrogen incorporation, affecting the $x$-value in $UH_{2+x}$. Since the H concentration influences both the 5$f$ electron localization and magnetic exchange interactions, even small deviations in stoichiometry may shift $T_C$. These observations stress the importance of substrate effects and growth conditions in controlling the magnetic behaviour of uranium hydride thin films.

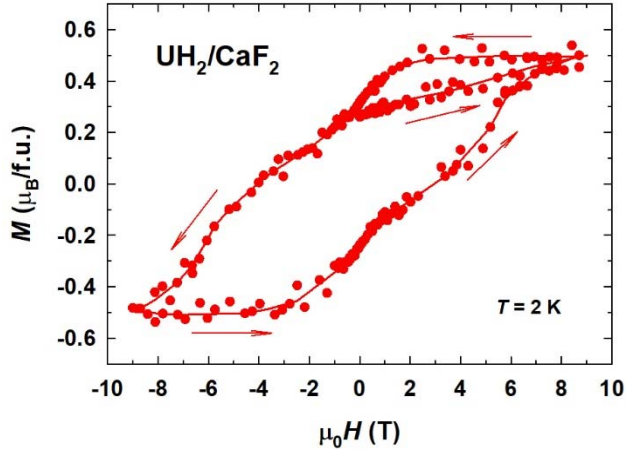 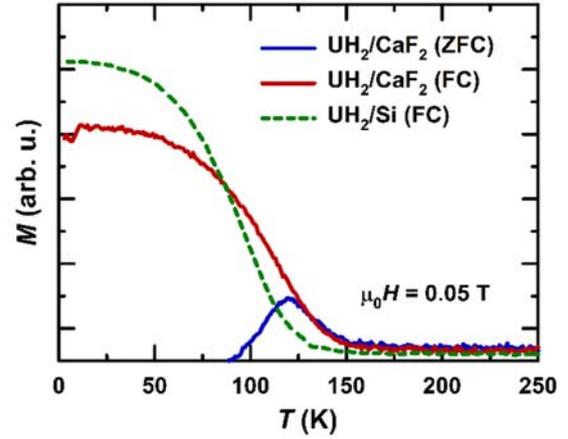

Fig. 6. The magnetic $M(H)$ hysteresis loop of the UH$_2$ film on the CaF$_2$ substrate in the magnetic field applied along the film plane at $T = 2$ K.

Fig. 7. ZFC and FC temperature dependencies of magnetization of UH$_2$ films measured in the field of 0.05 T applied along the samples' surface.

Comparing with magnetization measurement using the field applied perpendicular to the film surface (i.e., along the [1 1 1] crystallographic direction), we found that the signal is weak due to very limited amount of sample and shows no detectable hysteresis.

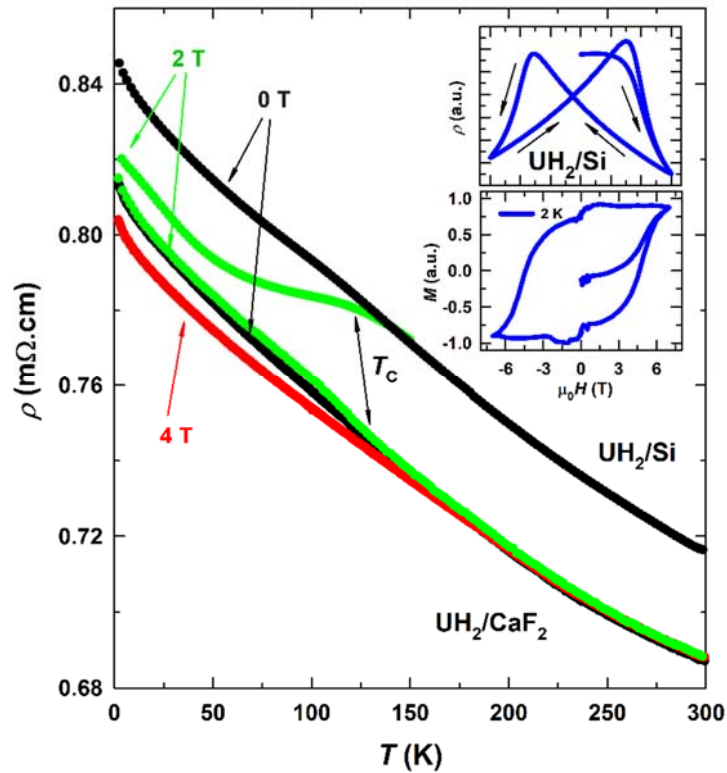

Fig. 8. Comparison of electrical resistivity $\rho(T)$ curves of UH$_2$ thin film samples grown on Si and CaF$_2$ substrates in various fields applied along the sample surface. The insets show the normalized magnetoresistivity and magnetization data at $T = 2$ K for the Si-based sample. The data for the UH$_2$/Si sample from [11] compared with UH$_2$/CaF$_2$ reflect that $T_C$ is indeed higher in the latter case but related anomaly is broader and the field effect weaker.

The temperature dependence of electrical resistivity (Fig. 8) exhibits for both films a negative slope (d$\rho$/d$T$ <0) in the whole temperature range. This is the effect of strong disorder and the fact that the continuation of the paramagnetic behaviour into the ferromagnetic state follows the same tendency means that the ferromagnetic ordering does not reduce the intense scattering of conduction. The increasing tendency with decreasing $T$ is somewhat attenuated by external magnetic field, which has been already observed in U hydrides. It can be taken as a sign of the fact that the dynamic disorder in the paramagnetic state continues in the form of a static disorder, the latter being affected by magnetic fields of several T [22,23]. Comparing the distinct sharp $T_C$ cusp in bulk nanocrystalline (U,T)H$_3$ hydrides [22], one can see a certain broadening for UH$_3$ films [11,23], even more broadening for UH$_2$/Si, and this trend continues to the UH$_2$/CaF$_2$ film. As all the systems have the grain size of few nm. These facts support the conjecture that the higher $T_C$ in the last film is associated with additional H sites occupied, yielding the UH$_{2+x}$ composition in the last case.

4.  **XMCD results and discussion**

XMCD studies were performed for both UH$_2$ films at 5 K in a magnetic field of 17 T applied parallel to the incident X-ray beam and at 15 degrees with respect to the sample surface. The estimated penetration depth of X-rays at the Uranium $M_{4,5}$ edges is about 300 nm, significantly exceeding the 40 nm thickness of the UH$_2$/CaF$_2$ film. As a result, for such a thin sample the effect of possible surface contamination, even with the Mo capping layer, becomes comparatively more pronounced. The normalized isotropic U-$M_{4,5}$-XAS spectra of 40-nm-thick UH$_2$ film on CaF$_2$ substrate are displayed in Fig. 9. The spectra are dominated by strong resonances, so-called white lines, originating from dipole-allowed transitions from the spin-orbit split 3$d$ core levels into the empty 5$f$ states, $3d_{3/2} \rightarrow 5f_{5/2}$ ($M_4$ edge) and $3d_{5/2} \rightarrow 5f_{5/2,7/2}$ ($M_5$ edge), along with much weaker $3d \rightarrow 6p$ and $3d \rightarrow$ continuum transitions. The $M_{4,5}$ XAS spectra represent thus a very sensitive probe of the occupancy of the 5$f$ levels that are quantified using branching ratio defined as $B=I_{M5}/(I_{M4}+I_{M5})$, where $I_{M5}$ and $I_{M4}$ are the integrated intensities of the white lines at the $M_5$ and $M_4$ edges, respectively. For this sample the value of $B$ is 0.697(2) which falls in between values 0.723 and 0.680 calculated within intermediate coupling scheme for pure ionic configurations $5f^3$ ($U^{3+}$) and $5f^2$ ($U^{4+}$), respectively [24].

This value of the branching ratio points to a non-integer $5f$ occupancy. If we assume a linear dependence of the branching ratio on the $5f$ count we would obtain the $5f$ occupancy of 2.4 in $UH_2/CaF_2$ and 2.7 in $UH_2/Si$. However, this approach relies on ionic calculations and therefore the delocalization of the $5f$ electrons could result in underestimation of the $5f$ count [25]. Conversely, any contribution from $UF_4/UO_2$ surface species would shift the $5f$ occupancy closer to 2, reflecting the $5f^2$ ground state of $U^{4+}$ in these ionic compounds [25].

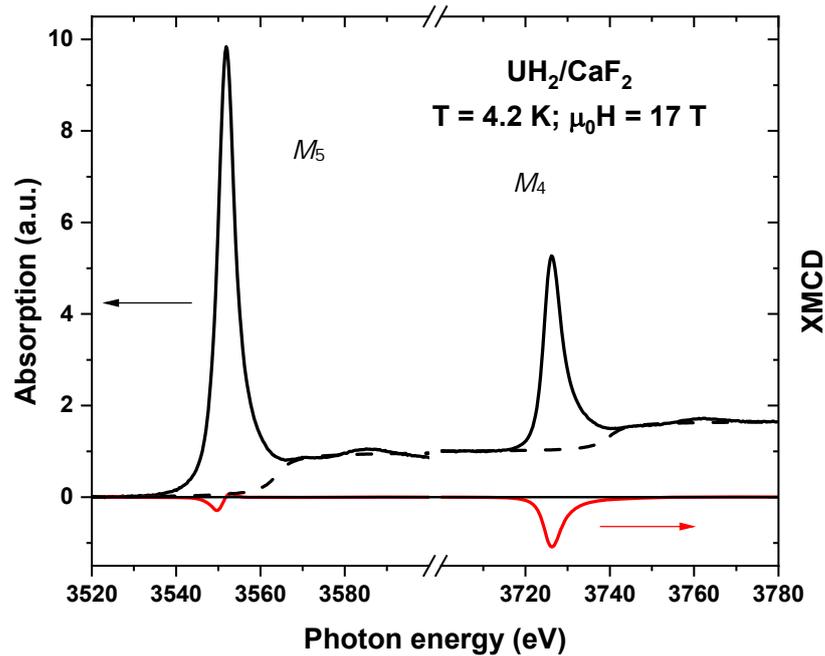

Fig. 9. X-ray absorption near edge structure (left axis, black curves) and x-ray magnetic circular dichroism (right axis red curves) spectra recorded at the U-$M_{4,5}$ edges for a $UH_2/CaF_2$ thin film sample in a magnetic field of 17 T applied in the grazing-incidence geometry (15° to the surface) at $T$ = 4.2 K as a function of the incident photon energy.

The XMCD spectra at the U-$M_{4,5}$ absorption edges reproduced on Fig. 9 (shown in red) have a typical shape for uranium-based intermetallics [15], i.e. a weak dispersive like signal at the $M_5$ edge and a more intense single asymmetric peak at the $M_4$ edge. To quantify the magnitudes of the magnetic moments carried by the uranium $5f$ states, we have made use of the so-called magneto-optical sum rules, which relate the integrals of the XMCD spectra ($\Delta I_{M_5}$ and $\Delta I_{M_4}$) to the orbital ($\mu_L^{5f}$=-$<L_z>\mu_B$) [26] and spin ($\mu_S^{5f}$=-2$<S_z>\mu_B$) [27] magnetic moments. Explicitly, the sum rules are given by

$$<L_z> = \frac{n_h^{5f}}{I_{M_5} + I_{M_4}}(\Delta I_{M_5} + \Delta I_{M_4})$$

$$<S_z> = \frac{n_h^{5f}}{2(I_{M_5}+I_{M_4})}\left(\Delta I_{M_5} - \frac{3}{2}\Delta I_{M_4}\right) - 3<T_z>.$$

The total U moment $\mu^U_{tot}$ (5f)= $-(<L_z> + 2 <S_z>)\mu_B$ can be determined from the XMCD spectra considering that the number of holes in the 5f shell ($n_h^{5f}=14-n_e^{5f}$) is equal to 11.6 as deduced from the branching ratio analysis of XANES spectra and the $<T_z>$ term is taken from atomic multiplet calculations [28].

Table 1. XMCD-derived spin ($\mu_S$), orbital ($\mu_L$), and total ($\mu_U = \mu_L + \mu_S$ moments for uranium in UH$_2$ thin films, compared with theoretical predictions [13]. The experimental uncertainty in the determination of the spin and orbital moments from XMCD spectra are estimated to be less than 10%.

| Sample | $n_e^{5f}$ | $\mu_S$ ($\mu_B$/U) | $\mu_L$ ($\mu_B$/U) | $\mu_{5f}^{total}$ ($\mu_B$/U) | $-\mu_L/\mu_S$ |
|---|---|---|---|---|---|
| UH$_2$/CaF$_2$ | 2 | −0.28 | 1.05 | 0.77 | 3.76 |
|  | 2.4 | −0.31 | 1.03 | 0.72 | 3.32 |
|  | 3 | −0.40 | 0.96 | 0.56 | 2.41 |
| UH$_2$/Si | 2 | −0.19 | 1.46 | 1.27 | 7.68 |
|  | 2.7 | −0.55 | 1.34 | 0.79 | 2.44 |
|  | 3 | −0.48 | 1.38 | 0.90 | 2.88 |
| Theory [13] | 2.8 | −2.05 | 2.94 | 0.89 | 1.43 |

As seen from Table 1, the resulting moments naturally depend on the 5f occupancy. The *BR* analysis described above provides a good estimate of the $n_{5f}$ value. Despite an apparent small difference (0.043) in BR values for $5f^2$ and $5f^3$ configurations, it represents >20% changes ratio of $I_{M5}/I_{M4}$ which is easily detectable. Surprisingly, this is often overlooked in the literature, referring to uncertainties in spectra normalization and background subtraction. Another useful indicator is the ratio of -$\mu_L/\mu_S$. Although some systems deviate strongly (such as nearly compensated UFe$_2$ [29] or other weak itinerant ferromagnets), a large number of U systems report -$\mu_L/\mu_S$ values in the range 2-3 [30, 15] depending on degree of localization of the 5f states. For the present UH$_2$ films, the deduced spin to orbital moments ratio would point out to $n_{5f} \approx 3$ for UH$_2$/CaF$_2$ and $\approx 2.7$-3 for UH$_2$/Si. The theoretical value 2.8 [13] is also consistent with this range.

The total moments obtained from XMCD are somewhat smaller for $UH_2/CaF_2$ (≈ 0.56-0.70 $\mu_B$) than for $UH_2/Si$ (0.70-0.90 $\mu_B$). As suggested above, a likely reason for the reduced magnitude in $UH_2/CaF_2$ is the presence of non-ferromagnetic, i.e. XMCD-silent, surface layer (e.g. $UO_2$ or $UF_4$), which is proportionally more significant for the thinner (~40 nm) film and may also slightly lower the *BR* values in XAS.

To place these experimental values in context, we compare them with available electronic-structure calculations. The theoretical data in Ref. [13] come from the GGA+*U* calculations (using the VASP package) with the PBE functional and rotationally invariant approach with $U = J = 0.5$ eV. Comparison with earlier electronic-structure approaches highlights the expected spread of spin and orbital moments depending on the computational scheme. Scalar relativistic LSDA calculations [10] yield spin moments 2.09 $\mu_B$/U. GGA ($U = 0$) gives $\mu_S$ even larger (2.19 $\mu_B$) while the total moments are much smaller (0.17 $\mu_B$). GGA+*U* with $U = 2.25$ eV [10] gives $\mu_S = 2.59$ $\mu_B$, total moment 0.45 $\mu_B$. This comparison suggests that there is a large spin moment without spin-orbit (s-o) coupling which then induces a large orbital moment when the s-o coupling is introduced. The calculated DOS in [13] also reveals the reason; 5*f* states are dominated by the spin-up states, with a large exchange splitting shifting the spin-down states above $E_F$. Within the present computational schemes, partial cancellation between the spin and orbital components leads to a total moment that is close to the experimental values (Table 1).

The spin and orbital moments from theory are about twice as high as the experimental values. A partial presence of unreacted (paramagnetic) U, which would reduce the values of XMCD-deduced moments, is unlikely, since such a mixture would yield a wasp-waisted hysteresis loop that is not observed (Fig 6 and inset of Fig. 8). Part of the discrepancy may instead arise because XMCD measures the field-aligned moment in textured films, which does not necessarily correspond to the easy-axis moment used in theoretical calculations. In addition, as shown above, the individual spin and orbital components are sensitive to the computational details.

**Concluding remarks:**

This study presents a comprehensive structural and magnetic characterization of uranium dihydride thin films, a metastable phase that cannot be obtained in bulk form. We focus on XMCD on $UH_2$ films comparing two films, one grown on Si, another on $CaF_2$(001) On $CaF_2$(001), the $UH_2$ film establishes a well-defined orientation relationship with the substrate and remains essentially strain-free; reciprocal space mapping yields a lattice parameter of 539 ± 3 pm, with any residual strain below the experimental resolution. By contrast, UH$_2$ films grown on Si with Mo buffer a polycrystalline fluorite-type structure in agreement with literature.

Magnetometry confirmed ferromagnetic ordering with $T_C \approx$ 120–130 K. A slightly higher $T_C$ for the CaF$_2$-supported film may reflect subtle differences in hydrogen stoichiometry. XMCD measurements provided a direct separation of orbital and spin components of the uranium 5$f$ magnetic moment in UH$_2$ thin films, but also the first quantification of U moments in the species, which has no bulk equivalent. The total moments below 0.9 $\mu_B$/U are slightly smaller than those from bulk magnetization studies of both UH$_3$ phases ($\approx$ 1.0 $\mu_B$) and much smaller than the microscopic moments of β-UD$_3$ (1.4 $\mu_B$ - see [14] and references therein). For the CaF$_2$-supported film, we obtain the U-5$f$ moment lower (below 0.7 $\mu_B$/U), which can be attributed to a surface U$^{4+}$ contribution (UO$_2$/UF$_4$), whose relative influence is enhanced due to the reduced thickness.

These results demonstrate that thin-film synthesis provides the viable route to stabilize UH$_2$ and to investigate its U magnetism by XMCD. Microgram amounts of natural or depleted U, far below the limit above which the radioactivity has to be taken into account, which opens access to large facilities without safety and security concerns. Moreover, assuming that the films can be easily accommodated into high pressure cells, it may open a way to synthesize U polyhydrides, predicted to exist under pressure and suspected to host high-$T_c$ [31].

**Acknowledgements:**

We acknowledge the support of Czech Science Foundation under the grant no. 22-19416S. The work of L.H. is supported by the grant 25-16339S. The samples were prepared in the framework of the EARL project of the European Commission Joint Research Centre, ITU Karlsruhe. Physical properties measurements were performed in the Materials Growth and Measurement Laboratory (http://mgml.eu/) supported within the program of Czech Research Infrastructures (Project No. LM2023065). Authors are indebted to D. Legut for illuminating discussion of the results of ab-initio computations.

**Supplementary information:**

## I. XAS/XMCD Measurement Details and Data Corrections

The XAS at both the U-$M_{4,5}$ edges were recorded in total fluorescence yield (TFY) detection mode in backscattering geometry, for parallel $\sigma+(E)$ and antiparallel $\sigma-(E)$ alignments of the photon helicity with respect to the external magnetic field (+/- 17 T) applied along the beam direction. The XAS spectra were then corrected for self-absorption effects. The corrected spectra are nearly identical to those recorded using the total electron yield (TEY) method, confirming the validity of the applied corrections (see Figure S1).

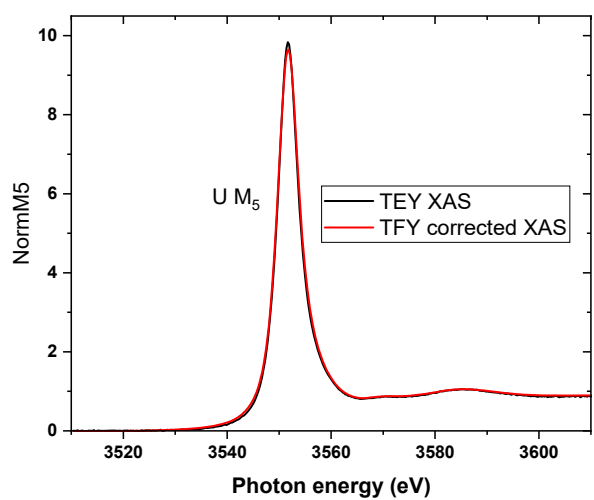

*Figure S1.* U-$M_5$ spectra measured with two techniques on the $UH_2/CaF_2$ sample.

## II. Geometric origin of the $UH_2$ lattice tilt on $CaF_2(001)$

To understand why $UH_2$ grows on $CaF_2(001)$ with a finite out-of-plane tilt rather than as a perfectly aligned (1 1 1) film, we examined the atomic registry between the fluorite lattices of the film and the substrate for different orientations. A purely (1 1 1)-oriented $UH_2$ film (i.e., with zero tilt) does **not** provide good atomic matching with the fourfold-symmetric $CaF_2(001)$ surface. Introducing a small tilt produces directions in which rows of U atoms align with rows in the substrate lattice, improving the interfacial registry. This is visible in the provided visualisations (Videos F1_unitcell_orientation_3_tilt.avi and _NOtilt.avi), where tilted configurations exhibit multiple azimuths with clear atom-on-atom "chains," while the untilted case does not.

We further visualized atomic arrangements for a continuous range of tilts between the (100) and tilted (1 1 1) orientations. For each tilt angle we extracted the atomic layer parallel to the surface (±0.5 Å slab) and compared its lateral registry with the CaF$_2$ surface atoms. The best match is obtained unsurprisingly for (0 0 1) orientation, however such orientation apparently does not grow within the used deposition conditions. We can only guess that under low-temperature conditions the growth is driven rather by the kinetics than achieving the most stable configuration. In that case growth close to [1 1 1] direction could be more preferred and the system would choose some orientation close to that, leastwise respecting the surface geometry. Although this is an approximate geometrical analysis—given the uncertainties in the exact lattice parameters and the fact that UH$_2$ distortion is below our experimental resolution—small tilt indeed improves local matching (Video F1_unitcell_orientation_surface_5.avi). The good lateral match is achieved for the tilt around 15°.

A key observation is that the (0 1 −1) planes in both CaF$_2$ and UH$_2$ remain parallel for any tilt angle, and their interplanar spacings are very similar. This provides a natural low-energy direction along which the interface can accommodate misfit.